# Wideband impedance measurement in the human ear canal; In vivo study on 32 subjects

Søren Jønsson, Andreas Schuhmacher, Henrik Ingerslev

Brüel & Kjær Sound & Vibration A/S, Skodsborgvej 307, DK-2850, Denmark

ABSTRACT

The human ear canal couples the external sound field to the eardrum and the solid parts of the middle ear. Therefore, knowledge of the acoustic impedance of the human ear is widely used in the industry to develop audio devices such as smartphones, headsets, and hearing aids. In this study acoustic impedance measurements in the human ear canal of 32 adult subjects is presented. Wideband measurement techniques developed specifically for this purpose enable impedance measurement to be obtained in the full audio band up to 20kHz. Full ear canal geometries of all subjects are also available from the first of its kind in vivo based magnetic resonance imaging study of the human outer ear. These ear canal geometries are used to obtain individual ear moulds of all subjects and to process the data. By utilizing a theoretical Webster's horn description, the measured impedance is propagated in each ear canal to a common theoretical reference plane across all subjects. At this plane the mean human impedance and standard deviation of the population is found. The results are further demographically divided by gender and age and compared to a widely used ear simulator (the IEC711 coupler).

# I. INTRODUCTION

In this study acoustic impedance measurements in the human ear canal of 32 adult subjects with normal hearing are presented. Impedance measurements of the human ear are only well covered in a limited bandwidth. From a literature review[1] no consistent date seems to be available above 8-10kHz. Here the wideband measurement technique developed in [1] is applied in order to obtain accurate and reliable impedance measurements in the full audio band up to 20kHz.

In the bare measurement results as presented a significant variability in impedance across subjects is noted, especially at frequencies above 3-4 kHz where resonances and anti-resonances occur. This variability is mainly due to a variable path length between the measurement planes in the ear canal and the ear drum, when measuring the impedance from subject to subject. It is difficult in practice to maintain the same path lengths, or insertion depth, across all subjects, partly due to the large intra-subject variation in ear canal geometries, where some ear canals are quite narrow with limited space to fit the waveguide tubes of the transmitter and receiver. It is also because the acoustical length of a human ear canal is not well defined. The acoustical length of a straight tube with uniform cross section is closely related to the physical length of the tube. For example, the first length resonance will occur at the frequency where a half wavelength becomes equal to the physical length of the tube. In a human ear canal the standing wave pattern is much more complicated, due to the curvature and difference in cross section area along the ear canal. Resonances and anti-resonances may therefore occur with different spacing in frequency than would be anticipated using a simple uniform tube approximation of the ear canal [2].

To compare features of ear canal impedances, and to obtain an estimate of the mean human ear canal impedance that preserves characteristic features of individual ear canal impedances, it is therefore desirable to propagate the impedance measurements to a common theoretical reference plane of the ear canal, so that resonances align up at the same frequency and level differences can be compared.

A uniform tube approximation of the ear canal has proven useful for propagating impedance measurements by means of the ear canal reflectance [3]. However, as the full audio frequency range up to 20kHz is of interest, a uniform tube approximation may not provide sufficient accuracy when propagating the impedances [4,5]. Webster's original horn equation [6] has often been applied to calculate the propagation of the sound field in straight one-dimensional waveguides with varying cross section area [7,4]. Here an extension to Webster's Horn equation shall be used [8,9] that also accommodate for the curvature of the ear canal, by predicting the longitudinal sound pressure along a curved axis following the centre of the ear canal. This approach is able to predict the sound pressure distribution reasonably well, also at higher frequencies, in the core region of the ear canal [10,11].

It does however require detailed knowledge of the ear canal geometry, such as the cross section areas normal to the centre line axis of the ear canal. This information is available from the first larger in-vivo magnetic resonance imaging (MRI) based study of the human outer ear [12]. The geometry of the ear canal from the bottom of the concha to the ear drum of 44 subjects has been accurately measured, 32 of which are the subjects included in this study. The main purpose of the MRI-based study was to calculate an average human ear canal while preserving distinct anatomical features, but those results are outside the scope of this paper. Here the individual ear canal geometries will first of all be used, to make individual ear moulds ensuring an optimal seal and fit when applying the impedance probe assembly to the ear of each subject. From the ear mould, the position of the measurements plane in the ear canal relative to the ear drum is now known and the required information is available to propagate the measurements along a centre line axis of the ear canal (back or forth) to a common reference plane.

Subsequently the measurements can be compared and an estimate of the mean human ear canal impedance and standard deviation may be calculated. Demographically division by gender and age is also presented.

Finally the mean human ear canal impedance is compared to an ear simulator, referred to as the IEC711 coupler, which is widely used to test various types of earphones and used as an integral part of a head and torso simulator, as defined in international standards and recommendations [13] [14] [15]. The results presented here is intended to form the basis for specifications of a wideband ear simulator, having an anatomical correct pinna and ear canal simulator, suitable for testing all types of earphones of smartphones, headsets, hearing aids, etc. in the full audio band up to 20kHz.

## II. METHODS

### A. The subjects

In total 32 adult subjects with normal hearing participated in this study. The demographics of the subjects are listed in

Table 1. The ages ranged from 26 to 60 years with a fairly even age distribution. Of the 32 subjects 19 were male and 13 were female, 23 right ear canals and 9 left ear canals, 25 from Denmark and 7 from the rest of the world.

Table 1: Demographics of the 32 adult subjects included in this study.

| Subject # | Gender | Age | Country of origin | Ear canal (R/L) | Subject # | Gender | Age | Country of origin | Ear canal (R/L) |
|---|---|---|---|---|---|---|---|---|---|
| **1** | M | 47 | DK | R | **21** | F | 43 | DK | R |
| **2** | M | 32 | DK | L | **22** | F | 41 | DK | R |

| 3  | F | 51 | DK | R | 24 | M | 51 | DK | R |
|----|---|----|----|---|----|---|----|----|---|
| 4  | F | 48 | DK | L | 25 | F | 40 | KR | L |
| 5  | F | 33 | DK | L | 26 | M | 41 | RU | L |
| 6  | M | 38 | DE | R | 27 | M | 51 | UK | L |
| 7  | M | 50 | DK | R | 28 | M | 45 | DK | R |
| 8  | F | 42 | DK | R | 31 | M | 38 | DK | R |
| 9  | F | 55 | DK | R | 33 | M | 41 | DK | R |
| 10 | F | 36 | LK | R | 34 | F | 33 | US | R |
| 11 | M | 50 | DK | L | 35 | M | 48 | DK | R |
| 15 | F | 60 | DK | R | 36 | M | 47 | DK | R |
| 16 | M | 46 | DK | R | 37 | F | 24 | DK | R |
| 17 | M | 44 | DK | R | 40 | M | 29 | DK | L |
| 19 | M | 42 | DK | L | 41 | F | 26 | DK | R |
| 20 | M | 30 | LK | R | 44 | M | 51 | DK | R |

## B. Measurement method and calibration

The wideband measurement technique developed in [1] is applied here in order to obtain accurate and reliable impedance measurements in the full audio band up to 20 kHz. The method is referred to as the semi multi-load calibration procedure. In short, as illustrated in Figure 1 transfer functions are measured on two known reference load cavities and then on an unknow load impedance, the ear canal of the test subject. To obtain a first estimate of the unknown load impedance these measurements are combined with known impedances for the two reference loads. This process is repeated for all test subjects over several sessions, meaning only a few subjects and transfer functions on the two reference loads are measured at each session. The semi multi-load calibration procedure is then performed as post processing to correct all estimated impedances. This procedure involves transfer functions to be measured only once on multiple reference loads, and moreover dividing the frequency range into two. The reference loads are simple cylindrical cavities, and their known impedances are defined by numerical simulations. For a complete review of the semi multi-load calibration procedure refer to [1].

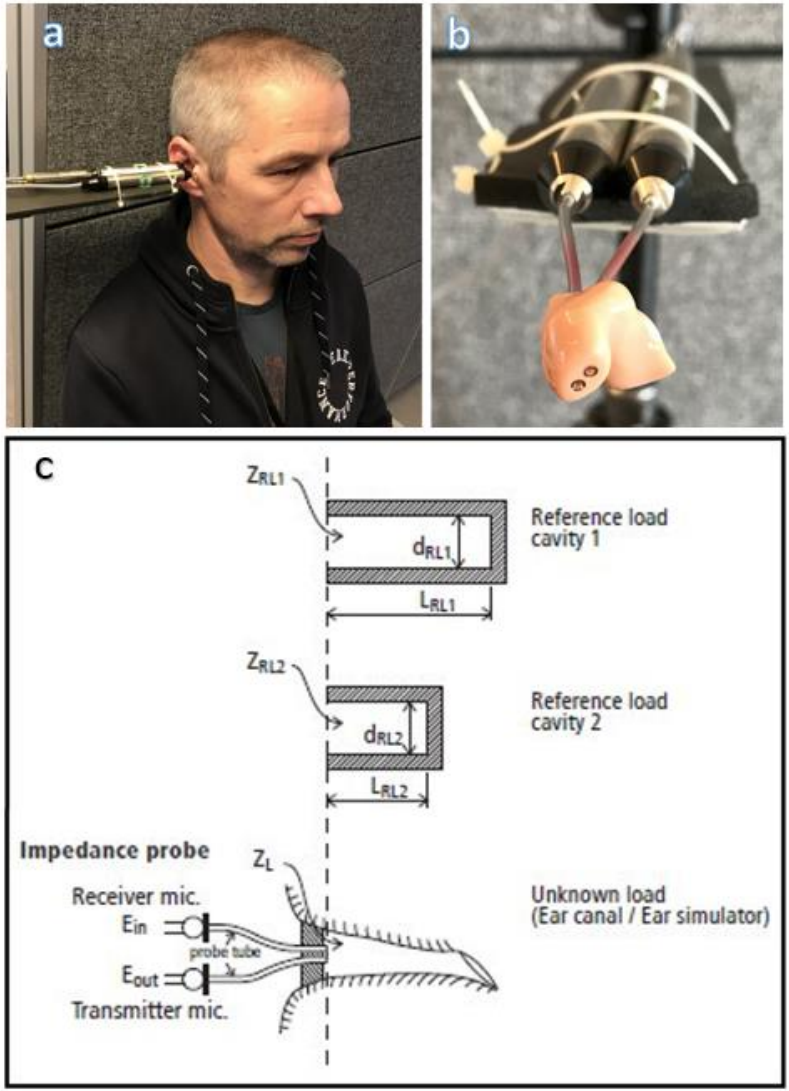

Figure 1 (a) A photo of the measurement setup with the impedance probe assembly inserted into the ear canal of one of the subjects. (b) A close-up of the impedance probe assembly with one of the individually fabricated ear moulds connected to the transmitter and receiver microphones through two thin acoustic waveguide tubes. (c) Schematic illustration of the impedance probe assembly with the receiver and the transmitter microphone on the left side which is in turn coupled to reference load cavity 1, reference load cavity 2, and the unknown load of the ear canal on the right side.

## C. Ear canal geometries and individual ear moulds

Detailed measurements of the ear canal geometry from the bottom of the concha to the ear drum of all subjects participating in this study is available from the first larger in vivo MRI-based study of the human outer ear [12]. In Figure 2 the geometry of the ear canals for subject 15, 26 and 27 are shown, which are examples of a small, medium and large ear canal, respectively. An overview of all ear canal geometries is included in Figure A1 in APPENDIX A.

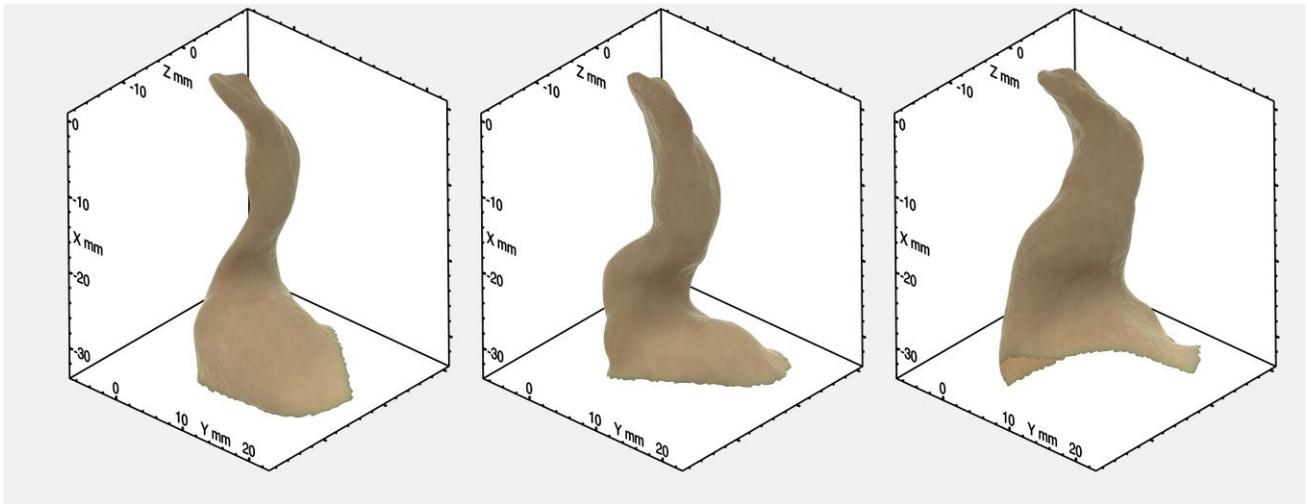

Figure 2: Magnetic resonance imaging measurement of the ear canal geometry of subjects 15, 26 and 27. They are examples of a small, medium and large size ear canal, respectively.

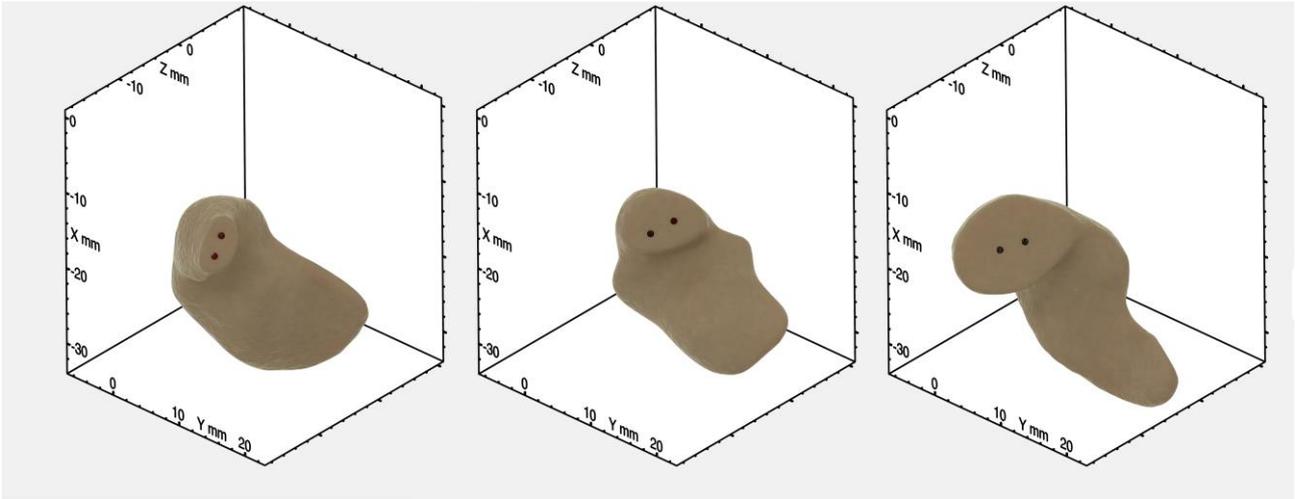

Figure 3: Images of three individual ear moulds for subjects 15, 26 and 27, as examples of a small, medium and large ear mould, respectively. The cross-sectional area at the measuring plane of the tip of ear mould is 17.5 mm$^2$, 46.0 mm$^2$ and 107.9 mm$^2$, respectively. The measuring plane of the ear mould is positioned between the first and second bend of the ear canal. The ear moulds have two holes for connecting the two acoustic waveguide tubes to the impedance probe assembly as shown in Figure 1 (b).

From the geometry data individual ear moulds for all subjects are fabricated. This ensures that an optimal seal and fit is obtained when inserting the impedance probe assembly into the ear canal, and that the measuring position in the ear canal is known, namely at the measurement plane of the ear mould (the tip of the ear mould). The measurement plane of the ear mould is positioned between first and second bend in the ear canal, and an attempt was made to have approximately the same insertion depth across all subjects. However, the path length from the measurement plane to the ear drum of the ear canal is difficult to keep the same across subjects, because of the large variation in ear canal geometries, and because the measurements session should not be too uncomfortable for the subject. Along a curved axis following the centre of the ear canal, the path length from the measurement plane of the ear mould to the bottom of the ear canal was between 17.3 mm and 25.3 mm, see

Table 2 column L1.

Figure 1 (b) shows a photo of one of the ear moulds with holes for two thin acoustic waveguide tubes. The holes are placed with the same spacing as used when connecting the impedance probe assembly to the reference load cavities as illustrated in Figure 1 (c). The acoustic wave guide tubes could easily be remounted in all the ear moulds, as well as reference load cavities, without introducing a leak. This allowed for the same set of tubes to be used throughout all measurement sessions. In Figure 3 individual ear canal moulds are shown for subjects 15, 26 and 27, as examples of a small, medium and large size ear canal with cross section area at the measurement plane of 17.5 mm$^2$, 46.9 mm$^2$ and 107.9 mm$^2$, respectively. An overview of all the ear moulds are shown in Figure A2 in APPENDIX A.

**D. Performing the measurements**

A day or two before a measurement took place, the subjects had their ear canals carefully cleaned with an agent and lukewarm water to ensure that any earwax or dirt was removed. If a subject was not well (had a cold or flu, etc.) on the scheduled day, the measurements was postponed. At the measurement session the ear mould was inserted in the ear canal of the subject with the two acoustic waveguide tubes already mounted. The subject was seated in a chair and the tubes were connected to the impedance probe assembly, see Figure 1 (a). A stepped sine sweep in 1/24$^{th}$-octave bands from 35Hz to 25kHz was then performed at a normal listening level. All measurements were repeated three or four times until the same overall frequency response could be obtained. It was ensured that no leakage was introduced between the ear mould and the ear canal, which usually appears as a roll-off at low frequencies. In a typical measurement session only three to four subjects were measured and it took about two months to complete the study. It is in general very difficult to obtain reliable measurements on human subjects even though careful preparations have been made. A total of 44 subjects entered into this study, but 12 were rejected because reliable measurements could not be obtained.

## III. THEORY

In this section it is explained how the measured ear canal impedances are propagated along a centre line axis of the ear canal to a common theoretical reference plane, so that the half wavelength resonance align at the same frequency. Webster's horn description of the ear canal is employed and a so-called Webster's horn propagation equation is derived. The procedure is illustrated by two examples of its use.

### A. Webster's horn propagation equation

Webster's horn equation [6] [16] is an approximation of the wave equation for a one-dimensional acoustic waveguide with rigid walls and variable cross section area, see Figure 4.

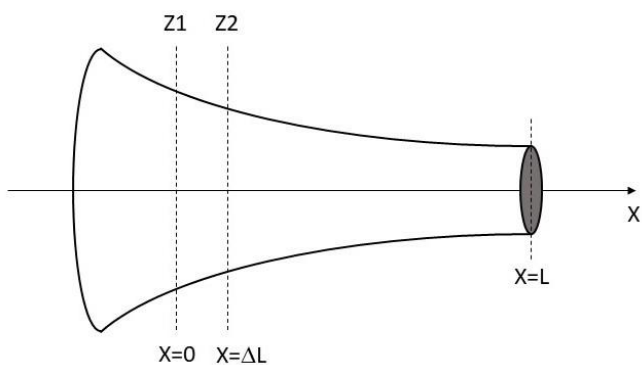

Figure 4: Schematic of the horn used to derive the propagation equation( 9 ). $Z1$ is the measured impedance at $x = 0$ (at the measurement plane of the ear mould) and $Z2$ is the propagated impedance at $x = \Delta L$ (the new theoretical measurement plane). The horn is terminated at $x = L$ with reflection coefficient $R$.

In velocity potential form it reads [17]:

$$\Phi'' + \frac{1}{A}A'\Phi' - \frac{1}{c^2}\ddot{\Phi} = 0 \qquad (1)$$

where $A$ is cross section area, $c$ is speed of sound and differentiation with respect to spatial coordinate $x$ and time $t$ is denoted with $'$ and $\cdot$ respectively. By using a time dependent solution of the form $\Phi = \Phi_1 \Phi_x(x) \exp(-i\omega t)$ with $k = \omega/c$, and substituting new variables; $A = \pi a^2$, where $a$ is the cross-section radius, and $\psi = \Phi_x \sqrt{A}$, equation ( b14 ) reduces to a one-dimensional Schrödinger equation [18] [19]:

$$\psi'' + \frac{a''}{a}\psi = -k^2\psi \qquad (2)$$

For a cone $a'' = 0$ and equation ( b16 ) reduces to:

$$\psi'' = -k^2\psi \qquad (3)$$

It has solutions of the form:

$$\psi = \psi_0 \exp(\pm ikx) \qquad (4)$$

The solution for the velocity potential for a forward moving wave and a backward moving wave reflected at $x = L$ with reflection coefficient $R$ is therefore given by:

$$\Phi = \frac{\Phi_0}{a}(\exp(ikx - i\omega t) + R\exp(-ik(x + 2L) - i\omega t)) \qquad (5)$$

Remember for a cone the cross-section radius $a = a(x)$ depends linearly on the spatial coordinate x.

The sound pressure is given by:

$$p = -\rho \frac{d\Phi}{dt} = i\rho\omega\Phi \qquad (6)$$

and the one-dimensional volume velocity $q = A\frac{d\Phi}{dx}$ is given by:

$$q = A\left(\frac{\Phi_0}{a}(ik\exp(ikx - i\omega t) - ikR\exp(-ik(x + 2L) - i\omega t)) - \frac{a'}{a}\Phi\right) \quad (7)$$

The acoustic input impedance can be found from equation ( 6 ) and ( 7 ),

$$Z = \frac{p}{q} = Z_0\left[\left(\frac{1 - R\exp(-2ik(x + L))}{1 + R\exp(-2ik(x + L))}\right) - \frac{ia'}{ak}\right]^{-1} \quad (8)$$

where $Z_0 = \rho c/A$ is the characteristic impedance.

If the impedance $Z1$ and cross-section radius $a1$ are measured at $x = 0$, the impedance $Z2$ at $x = \Delta L$ where the cross-section radius is $a2$ (see Figure 4) can be found by setting up equation( b24 ) in these two cross-sections and then eliminate the reflection $R$ and solve for $Z2$, which yields,

$$Z2 = Z2_0\left[\frac{\left(Z1_0 + i\frac{Z1a1'}{ka1}\right) - Z1 i\tan(k\Delta L)}{Z1 - \left(Z1_0 + i\frac{Z1a1'}{ka1}\right)i\tan(k\Delta L)} - \frac{ia2'}{ka2}\right]^{-1} \quad (9)$$

where $Z1_0$ and $Z2_0$ are the characteristic impedance at the two cross-sections. In the following equation ( 9 ) will be referred to as the Webster's horn propagation equation. For a full derivation of equation ( 10 ) see APPENDIX B.

For a cylindrical segment we have $Z1_0 = Z2_0$, and $a1' = a2' = 0$, and equation ( 9 ) reduces to the propagation equation for cylinders [20][21]:

$$Z2 = Z1_0\frac{Z1 - Z1_0 i\tan(k\Delta L)}{Z1_0 - Z1 i\tan(k\Delta L)} \quad (11)$$

By slicing the geometry of the human ear canal into a large number of sections along a curved axis following the centre of the ear canal, as illustrated in Figure 5, and by approximating each of those sections with a cone, Webster's horn propagation equation ( 12 ) can be used in each section to find the impedance at the end plane (Z2) of the cone, knowing the impedance at the start plane (Z1) of the cone [8] [9].

## B. The centre line

Cross-sectional areas of the ear canal must be extracted normal to a curved axis following the centre of the ear canal. These centre lines are obtained from the surface model of all the ear canal geometries. The centre line is a line down through the ear canal that trace out its centre. More specifically, it is calculated from the measured ear canal geometries by stepping a sphere down through the ear canal while maintaining the maximum radius of the sphere that inscribes the ear canal at each step. The centre line is then the path that the centre of this sphere is stepped along. [22] [23]. In Figure 5(a) the centre line of subject 4 is shown.

## C. Two examples of propagating the impedance

The geometry files of the ear canal and of the ear moulds are obtained from the imaging (MRI) based study of the human outer ear [12].

The input impedance in the ear canal of each test subject is measured at measurement plane of the ear mould (the tip of the ear mould). To find the corresponding plane in the geometry of the ear canal, the geometry of the ear mould is aligned inside the geometry of the ear canal by minimizing the distance between the wall of the ear mould and the wall of the ear canal. Figure 5(a) shows an ear canal (blue) and an ear mould (red) after such an alignment. There are also two green intersection planes, where the first plane is parallel to the measurement plane of the ear mould, and the second is a new theoretical reference plane where the propagated ear canal impedance can be calculated using Webster's horn propagation equation( 9 ). The section of the ear canal in between the two intersection planes is further intersected into many closely spaced planes with a separation of about 0.05 mm, all perpendicular to the centre line as illustrated in Figure 5(c). The impedance at plane $n + 1$ can be found from plane $n$ using Webster's horn propagation equation. Hence, the impedance at the end-plane ($ZN$) can be found by using Webster's horn propagation equation $N - 1$ times, where $N$ is the number of intersection planes. The result of this procedure is shown in

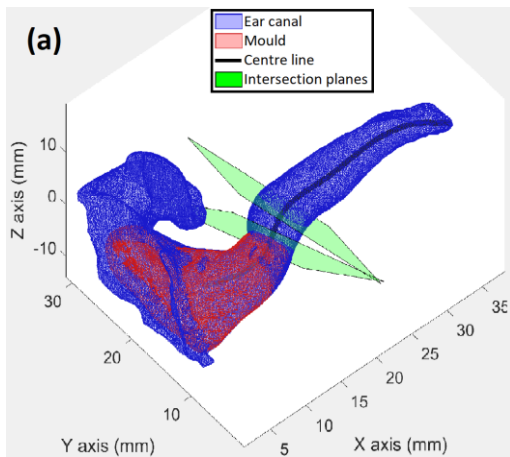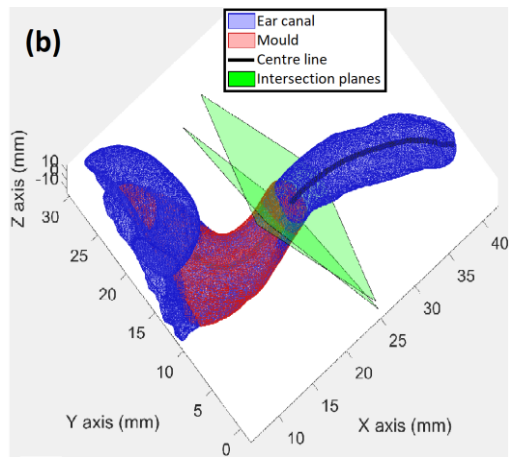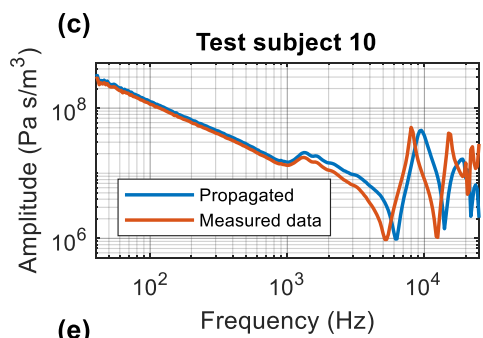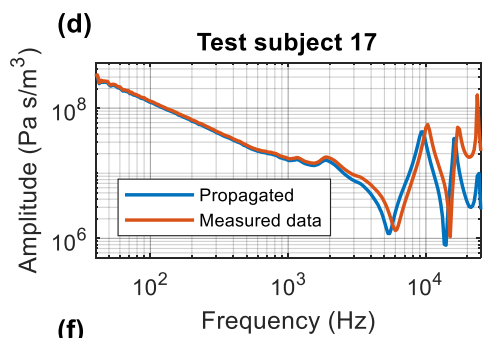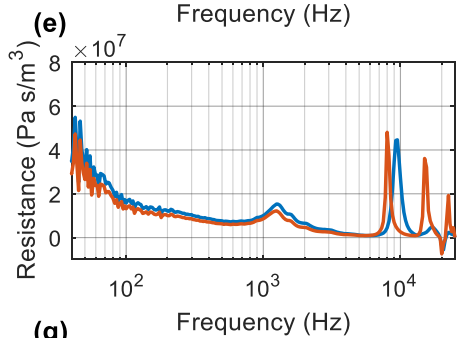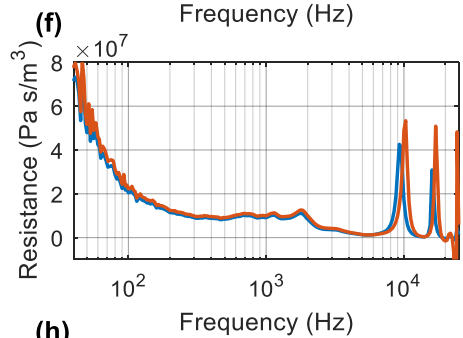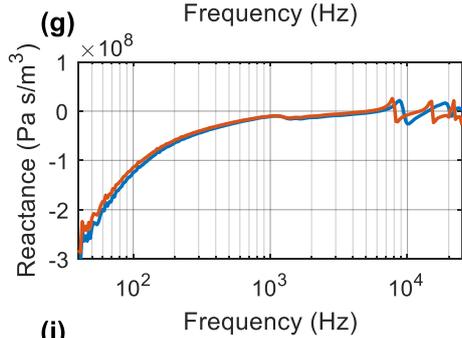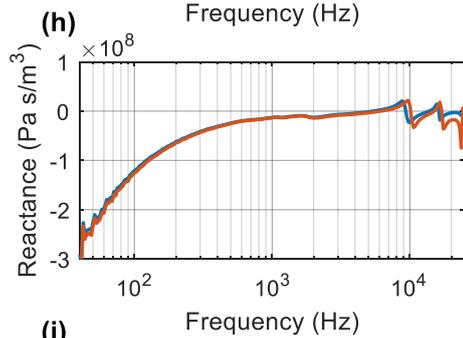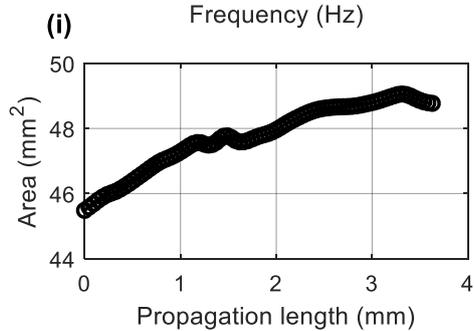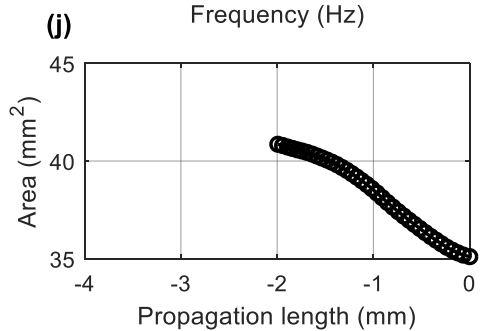

Figure 6 for test subjects 10 and 17, where the measurement plane is moved +3.6 mm and -2.0 mm respectively resulting in the half-wavelength resonance being shifted to align at 9.40kHz. Similar propagations are made for all subjects so that the half-wavelength resonance is shifted to align at 9.40kHz as shown in Figure 7 for the first 5 subjects, and in Figure A1 in the APPENDIX A for all 32 test subjects.

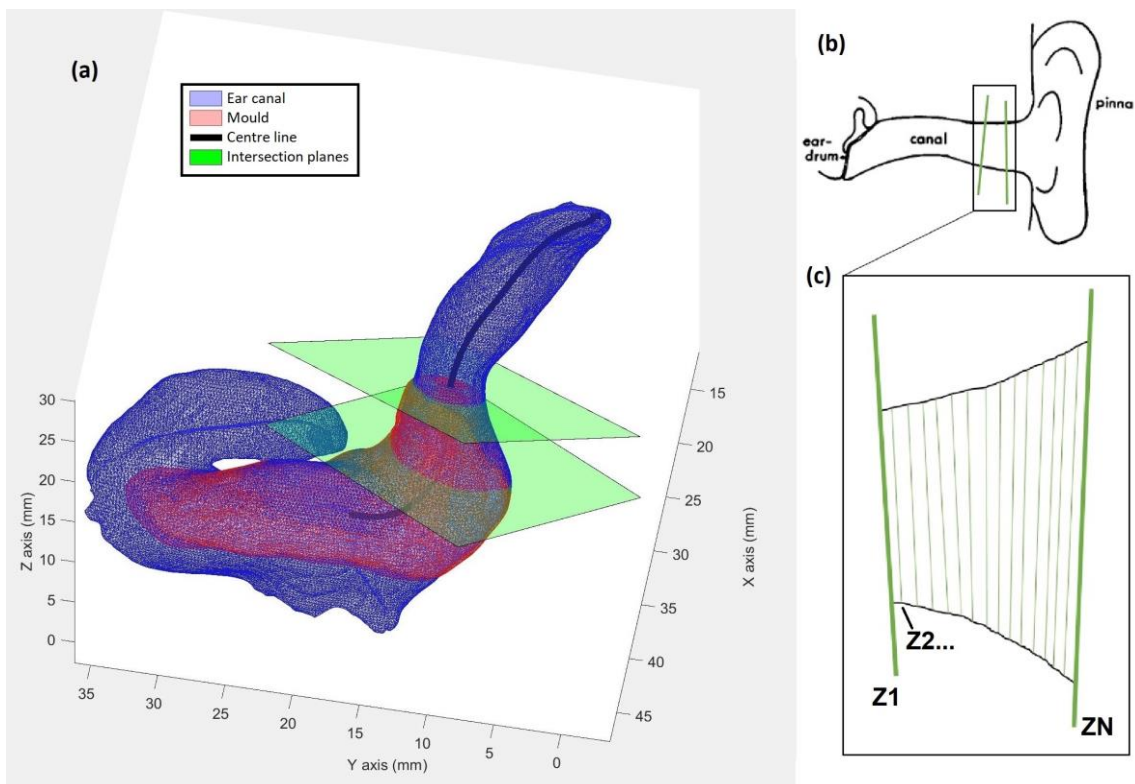

Figure 5: (a) Geometry of the ear canal and an ear mould for test subject 4. The ear mould is aligned inside the ear canal by minimizing the wall-to-wall distance. The centre line and two intersection planes are also shown. The first intersection plane is at the measurement plane of the ear mould closest to the ear drum. The second plane is the new reference plane where the half wavelength resonance of the propagated impedance will align at 9.40 kHz, when using the procedure described in the text. Between the two intersection planes shown in (a) there are multiple closely spaced planes, as illustrated in (b) and (c), used for calculating the propagated ear canal impedance.

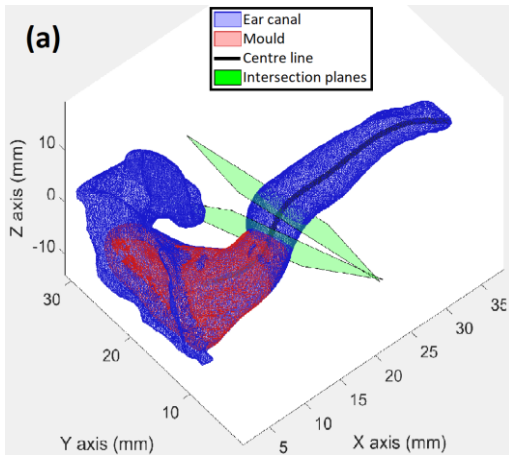 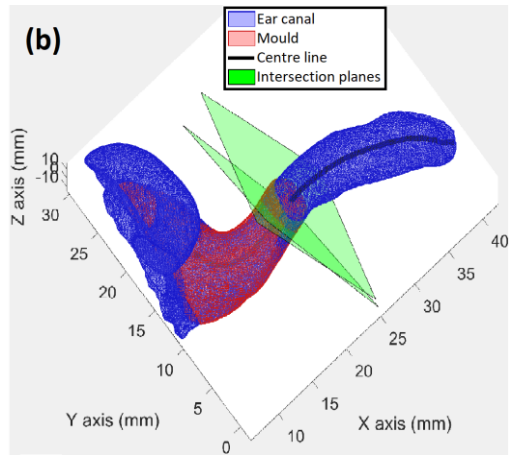
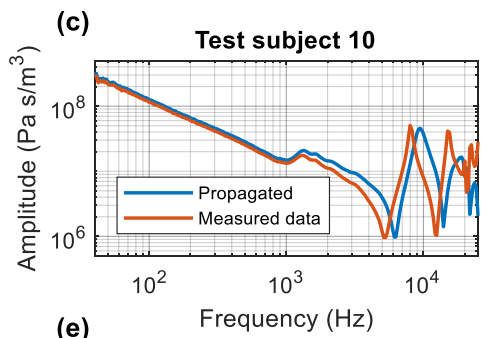 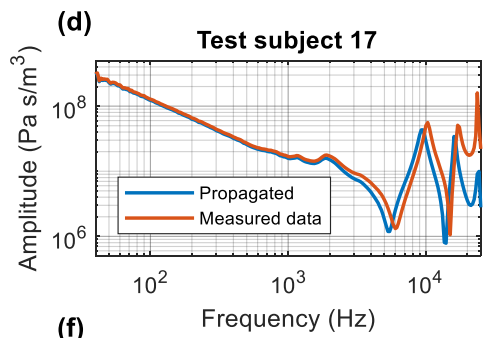
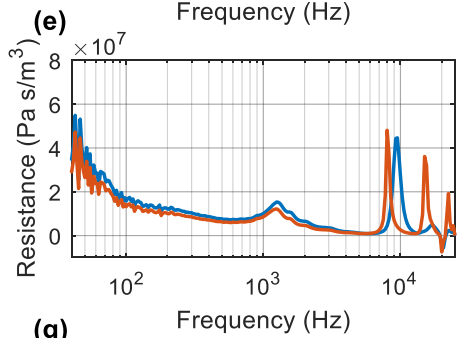 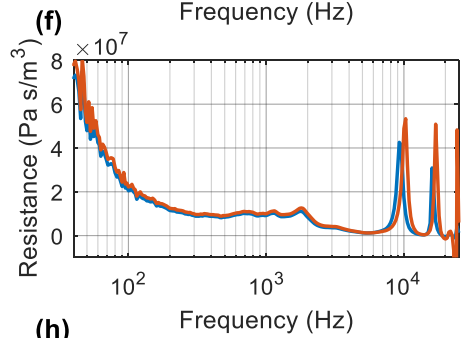
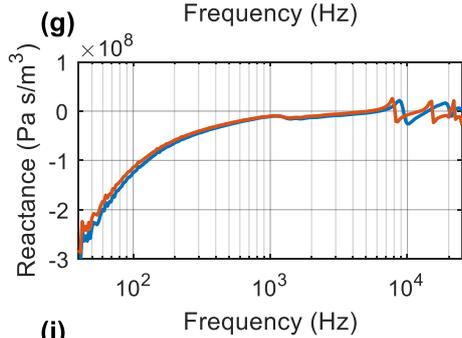 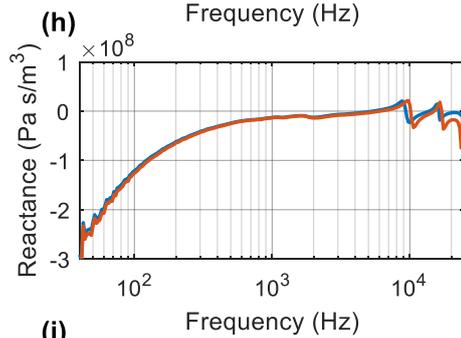
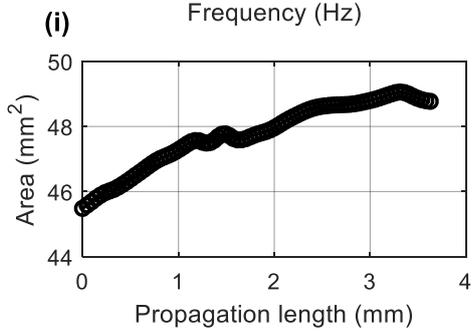 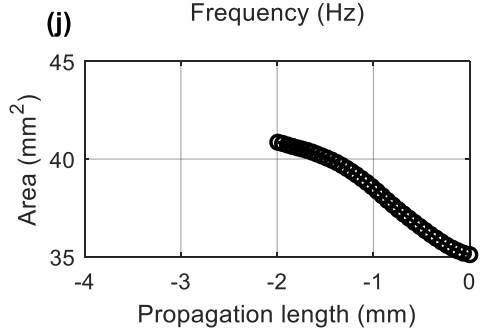

Figure 6: Two examples of propagating the measured ear canal impedance of test subject 10 and 17, resulting in an upward and downward shift (respectively) of the half wavelength resonance so they align at 9.40 kHz. (a) and (b) shows the geometry of the ear canal and mould, and the start and end cross section plane. (c) to (h) show amplitude, real and imaginary part of the measured and propagated ear canal impedance. (i) and (j) show the cross section area at each of the many closely spaced planes between the start and end cross section plane.

## IV. RESULTS

The results are presented in the following five sections. First, the bare calibrated impedance measurements at the measurement plane are presented. Next, the propagated impedance at the common theoretical reference plane are presented. Then the mean and standard deviation of the propagated impedances are presented. The propagated impedances are next demographically divided based on gender and age, and finally the mean propagated impedance are compared to a the propagated impedance of a widely used ear simulator, the IEC711 coupler.

### A. Impedance at the measurement plane

In the left column of Figure 7 and Figure A3 in the APPENDIX Ax, the impedance at the measurement plane of the ear mould of the first five subjects and all 32 subjects are shown, respectively. The impedance is calibrated with the semi multi-load calibration procedure as described in section II.B. The upper graph shows the impedance magnitude and the two lower graphs shows the real and imaginary part of the impedance, i.e. the resistance and reactance.

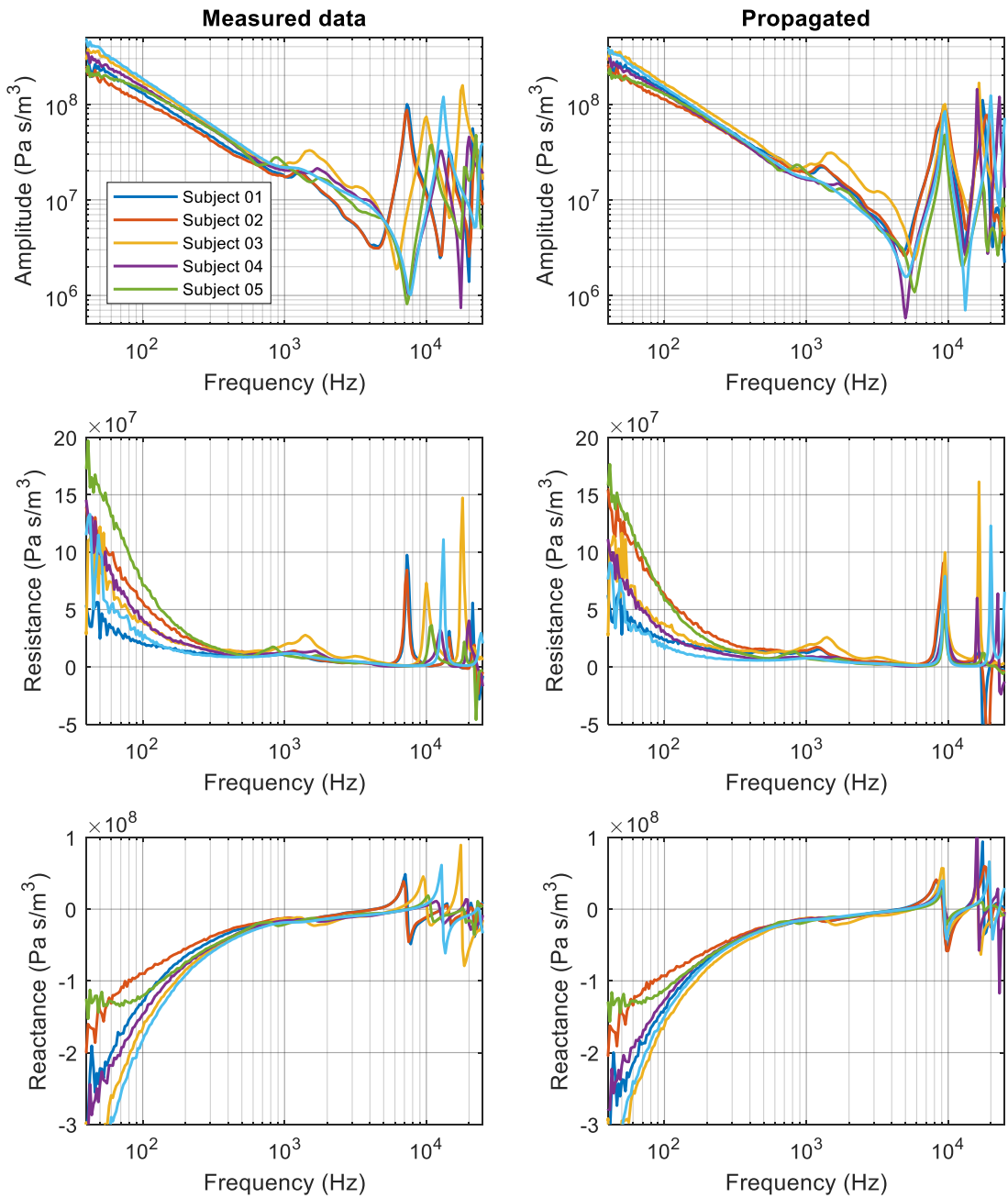

Figure 7: Measured (left column) and propagated impedance (right column) of five adult subjects (01, 02, 03, 04, 05). The upper graphs show the amplitude and the two lower graphs show the real and imaginary part, i.e. the impedance resistance and reactance respectively.

## B. Propagated impedance

All measurements are propagated as described in section III along the centre line of the ear canal (back or forth) to a common reference plane, so that resonances at higher frequencies align and individual level differences can be compared. More precisely, the measurements are propagated, so that the half-wavelength resonances align at 9.40 kHz. The frequency of 9.40 kHz is selected the following way. The average calculated wavelength at the half wavelength resonance frequency for all 32 subjects gives 36.5 mm, which corresponds to a frequency of 9.40 kHz. Along the curved centre line of the ear canal the average path length from the reference plane to the bottom of the ear canal of is 22 mm. See

Table 2 which contains an overview of the propagation lengths, the cross-sectional areas, and the half-wavelength resonances at the measuring plane, and at the propagated plane for all subjects, and for the artificial ear, the IEC711 coupler. Mean and standard deviation of each parameter for the 32 subjects are also given.

In the right column of Figure 7 and Figure A3 in the appendix, the propagated impedances are presented for the first five subjects and for all 32 subjects, respectively. The upper graph shows the impedance magnitude and the two lower graphs shows the impedance resistance and reactance.

Table 2: Overview of the propagated path length for all subjects and of an artificial ear, the IEC711 coupler. A1 and A2 are the cross-sectional areas of the ear canal at the measuring plane and at the propagated plane. L1, L2 are the path lengths from the measuring plane to the bottom of the ear canal along the centre line of the ear canal before and after propagation, respectively. $\Delta L$ is the propagation length. And f1 and f2 are the half-wavelength resonance before and after propagation. The rows named Mean and STD are the mean and standard deviation of the 32 subjects.

| Subject number | A1 (mm²) | A2 (mm²) | L1 (mm) | L2 (mm) | $\Delta L$ (mm) | f1 (kHz) | f2 (kHz) |
|---|---|---|---|---|---|---|---|
| 1 | 64.4 | 65.7 | 24.2 | 22.1 | 2.08 | 7.35 | 9.42 |
| 2 | 45.1 | 62.2 | 22.8 | 20.0 | 2.81 | 7.26 | 9.45 |
| 3 | 27.0 | 26.7 | 20.6 | 22.0 | -1.33 | 9.95 | 9.40 |
| 4 | 33.5 | 60.2 | 19.8 | 25.9 | -6.14 | 12.66 | 9.36 |
| 5 | 53.5 | 50.0 | 19.0 | 23.2 | -4.18 | 10.77 | 9.39 |
| 6 | 38.5 | 44.4 | 22.3 | 19.5 | 2.79 | 8.70 | 9.42 |
| 7 | 52.2 | 81.8 | 19.6 | 23.6 | -4.00 | 10.97 | 9.40 |
| 8 | 51.5 | 49.2 | 20.9 | 22.4 | -1.49 | 10.31 | 9.36 |
| 9 | 43.4 | 41.3 | 20.7 | 21.4 | -0.80 | 9.61 | 9.39 |
| 10 | 45.5 | 48.8 | 25.3 | 21.7 | 3.63 | 8.05 | 9.41 |
| 11 | 32.7 | 33.6 | 22.9 | 22.1 | 0.79 | 9.08 | 9.42 |
| 15 | 17.5 | 24.1 | 19.7 | 23.1 | -3.41 | 11.25 | 9.38 |
| 16 | 56.0 | 47.5 | 24.8 | 22.3 | 2.56 | 7.37 | 9.44 |
| 17 | 35.1 | 40.9 | 21.1 | 23.1 | -1.99 | 10.26 | 9.38 |
| 19 | 38.3 | 42.1 | 20.1 | 25.6 | -5.53 | 11.18 | 9.39 |
| 20 | 36.1 | 43.4 | 22.7 | 20.6 | 2.10 | 8.96 | 9.41 |
| 21 | 42.5 | 46.0 | 20.1 | 21.8 | -1.77 | 10.27 | 9.37 |
| 22 | 32.8 | 32.6 | 20.8 | 21.0 | -0.23 | 9.48 | 9.40 |
| 24 | 22.0 | 21.3 | 20.3 | 21.5 | -1.24 | 9.89 | 9.39 |
| 25 | 35.9 | 36.0 | 24.3 | 24.3 | -0.05 | 9.41 | 9.34 |
| 26 | 46.0 | 40.9 | 24.9 | 23.1 | 1.85 | 8.85 | 9.41 |
| 27 | 107.8 | 103.3 | 23.1 | 22.4 | 0.73 | 8.98 | 9.45 |
| 28 | 36.1 | 37.4 | 22.6 | 23.8 | -1.21 | 9.82 | 9.34 |
| 31 | 51.1 | 50.5 | 20.5 | 21.1 | -0.58 | 10.05 | 9.35 |
| 33 | 52.6 | 53.1 | 23.0 | 22.5 | 0.43 | 9.25 | 9.42 |
| 34 | 40.6 | 30.2 | 17.3 | 22.7 | -5.38 | 11.71 | 9.38 |
| 35 | 55.7 | 54.6 | 19.8 | 19.6 | 0.20 | 9.30 | 9.42 |
| 36 | 86.1 | 84.7 | 21.3 | 21.1 | 0.25 | 9.27 | 9.41 |
| 37 | 47.2 | 45.9 | 19.7 | 19.0 | 0.71 | 9.14 | 9.43 |
| 40 | 67.0 | 66.0 | 22.9 | 21.6 | 1.31 | 8.62 | 9.41 |
| 41 | 54.2 | 52.9 | 21.2 | 21.0 | 0.24 | 9.26 | 9.41 |
| 44 | 64.5 | 70.3 | 21.8 | 19.9 | 1.95 | 8.77 | 9.40 |
| Mean | 47.3 | 49.6 | 21.6 | 22.0 | -0.47 | 9.56 | 9.40 |
| STD | 17.8 | 18.0 | 1.9 | 1.6 | 2.58 | 1.23 | 0.03 |

| IEC711 | 44,2 | 44.2 | 12.5 | 19.0 | -6.50 | 13.20 | 9.42 |

## C. Mean human ear canal impedance HERTIL

In Figure 8 the mean and standard deviation of all 32 propagated impedances are presented. The mean human ear impedance is very well determined from the lowest frequencies and up to 4-5 kHz, where the standard deviation is only about 20-30% of mean. At frequencies from 4-5 kHz up to 13-14 kHz the impedance is fairly well determined, since the standard deviation raises to about 50-90% of mean and only slightly above 100% at the quarter wavelength resonance. Beyond 13-14 kHz the impedance is poorly determined since the standard deviation here generally is higher than mean.

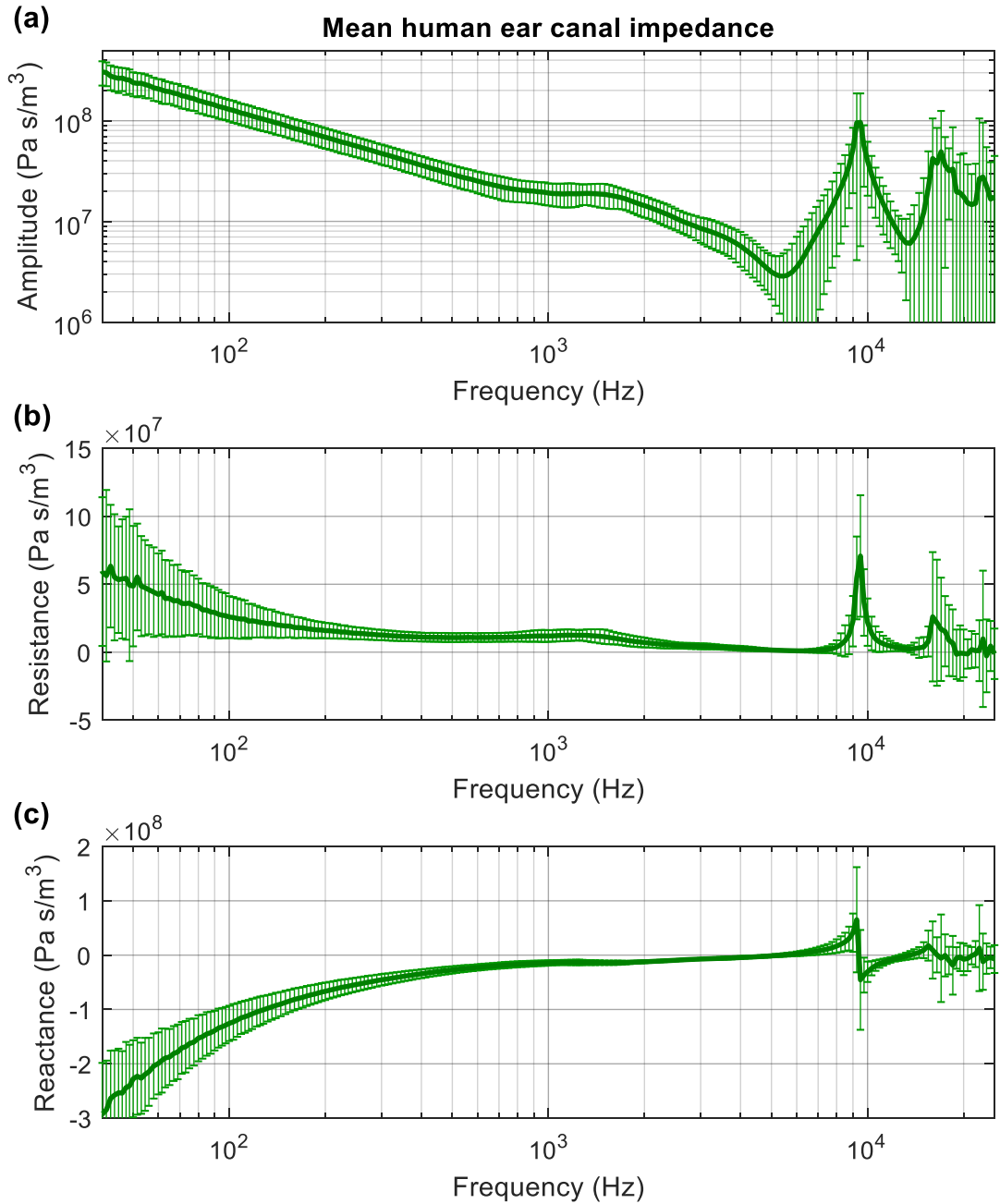

Figure 8: Mean human ear canal impedance of all 32 subjects, with error bars of one standard deviation, and where each of the 32 impedance measurements are propagated, so the half wavelength resonance aligns at 9.4kHz across all subjects. (a), (b) and (c) shows the amplitude, real part, and imaginary part respectively.

**D. Demographic division**

The 32 propagated impedance measurements are demographically divided into female and male groups, and for each group the mean and standard deviation is calculated as shown in Figure 9(a). Similarly, demographical division into two age groups of young adults from 26 to 42 years, and middle-aged adults from 43 to 60 years are shown in Figure 9(b). The mean of the female group is within one standard deviation of the male group, and vice versa. The mean of the young adult group is within one standard deviation of the middle-aged adults, and vice versa. Hence, no statistical significant difference in ear canal impedance between female and male, and young adults and middle-aged adults can be concluded based on the 32 subjects in this study.

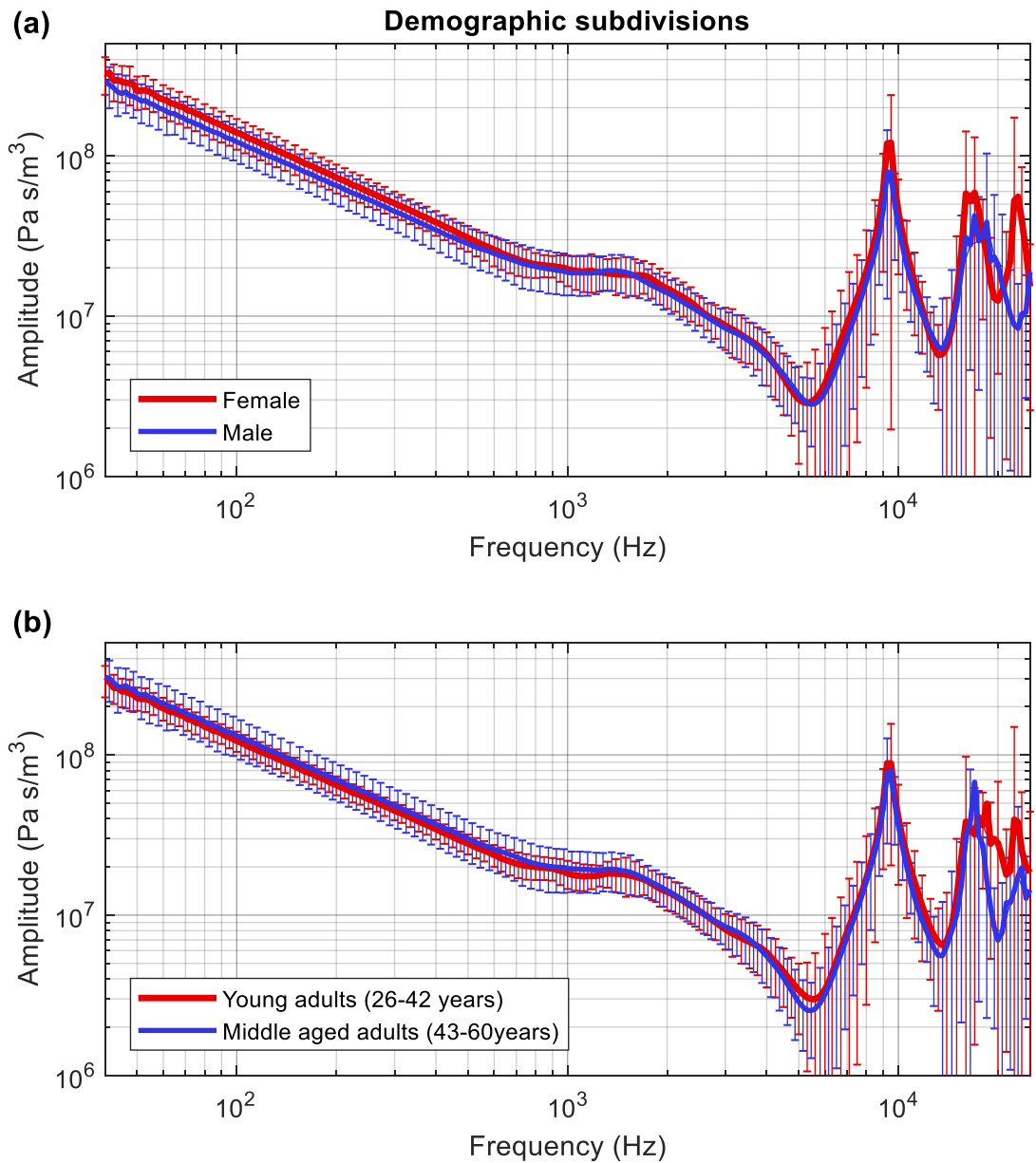

Figure 9: Demographical division of the propagated ear canal impedance of the 32 subjects. (a) Division into female and male. (b) Division into young adults from 26 to 42 years, and middle-aged adults from 43 to 60 years. Each graph shows mean with error bars of one stand deviation.

## E. Comparison with an ear simulator (the IEC711 coupler)

Here the mean human ear canal impedance is compared to an ear simulator as defined in the international standard IEC60318-4[13]. Formerly it was named the IEC711 standard and it is therefore often referred to as the IEC711 coupler. It is widely used for hearing aid testing of earphones coupled to the ear by ear inserts, such as tubes or ear moulds and also widely used as an integral part of all Type 3 ear simulators for measurements on telephones and headsets as specified in ITU-T Rec. P.57[14] and ITU-T Rec. P.58[15]. ITU-T Rec. P.58 specifies a head and torso simulator for telephonometric measurements. In this configuration, which will be used here, a cylindrical ear canal extension is added to the IEC711 coupler. The impedance of the IEC711 coupler has only been verified in the frequency range from 100 Hz to 8 kHz. Beyond that frequency range it is unknown how well it correlates with the human ear, even though it is widely used for testing of all types of earphone devices.

The measured impedance of the IEC711 coupler is shown together with the mean and standard deviation of the human ear impedance in Figure 10. The IEC711 impedance is also propagated, in its cylindric ear canal extension, so the half wavelength resonance align at 9.40 kHz, see

Table 2. In general, the impedance of the IEC711 coupler has the same overall form as the mean human ear. However, there are some noticeable differences. At the lowest frequencies the level of the IEC711 coupler is slightly higher, probably indicating a slightly too low equivalent volume of the IEC711 coupler. It is also noted that the slope of the curve towards mid-frequencies follows a -6 dB pr. octave, whereas the slope of the mean human impedance ear is slightly less steep. This indicates that the human impedance cannot be approximated by a simple compliance towards the lowest frequencies. From 1 kHz to 5 kHz the IEC711 coupler underestimates the mean human ear by about 40% (3dB). Around the half-wavelength resonance at 9.40 kHz the two curves match quite well, whereas around the quarter wavelength anti-resonance at 5-6 kHz the two curves match poorly. This is in part because both the IEC711 coupler and the human ear canals are shifted to 9.40 kHz, and in part probably because the simple cylindric ear canal shape of the IEC711 coupler is not quite able to reproduce the complex standing wave pattern of the human ear canals, which is also noted above 9.40 kHz where til anti-resonance and resonance of the IEC711 coupler are quite pronounced. At the three-quarter wavelength anti-resonance around 15 kHz and beyond, the two curves match poorly, but in this region the standard deviation of the human ear impedance is also relatively high, which makes a comparison difficult.

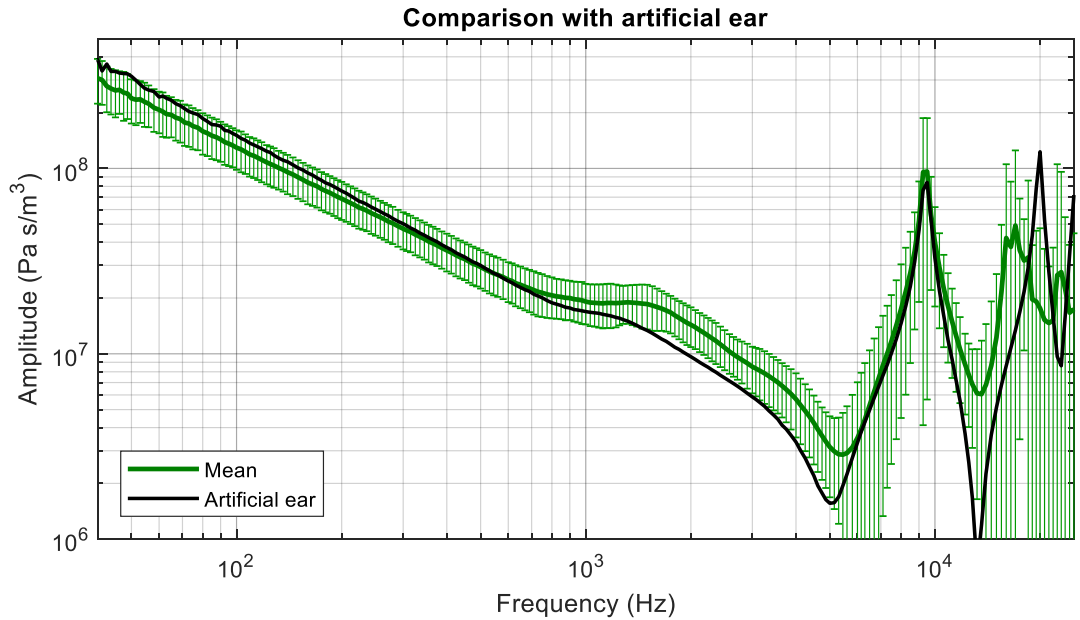

Figure 10: Comparison of the mean human ear impedance with error bars of one standard deviation to an ear simulator, the IEC711 coupler.

## V. CONCLUSION

In this study impedance measurements in the ear canal of 32 human subjects were presented.

Impedance measurements of the human ear are only well covered in a limited bandwidth below 8-10 kHz. Several factors contribute to difficulties in obtaining accurate and reliable measurements, such as standing waves at higher frequencies where wavelength of sound becomes close to the dimensions within the ear canal, that measurements traditionally are obtained in the ear canal at some unknown distance from the ear drum, issues with leakage introduced between the ear mould and the ear canal, the condition of the subject on the day the measurement is scheduled to take place. Here, the semi multi-load calibration procedure, a wideband measurement technique developed for the purpose, was applied in order to obtain accurate and reliable impedance measurements in the full audio band up to 20kHz.

Next, the impedance measurements were propagated to a theoretical common reference plane in the ear canal ensuring that the half wavelength resonance was aligned at the same frequency, 9.4kHz across all subjects. This was achieved using an extended Webster's horn description of the ear canal, which required full ear canal geometries of all subjects. These were available from the first of its kind larger in-vivo magnetic resonance imaging (MRI) based study of the human outer ear.

At the theoretical reference plane of the ear canal the mean human impedances and standard deviation was calculated. The mean impedance was well determined in most of the frequency range up to around 15 kHz with a standard deviation smaller than the mean, except around a small region at the quarter wavelength resonance.

Demographical division into female and male, and young adults (26 to 43 years) and middle ages adults (44 to 60 years) was also presented and showed no significant statistical difference.

Finally, the mean human ear canal impedance was compared to an ear simulator, the IEC711 coupler, which is widely used to test various types of earphones and used as an integral part of a head and torso simulator, as defined in international standards and recommendations within IEC (electroacoustics) and ITU-T (telecommunications). The same overall impedance response could be recognized, however with noticeable differences. At lower frequencies the level of the IEC711 coupler was slightly higher, and in the midfrequency area it underestimates the mean human with just over one standard deviation, and at higher frequencies above the half wavelength resonance the standing wave pattern was quite different.

The results presented will form the basis for specifications of a wideband ear simulator, having an anatomical correct pinna and ear canal simulator, suitable for testing all types of earphones of smartphones, headsets, hearing aids, etc. in the full audio band up to 20kHz.

## VI. ACKNOWLEDGEMENT

We acknowledge the time and effort from all the subjects participating in this study, and for suggestions and comments from the acoustic transducer team at Brüel & Kjær.

# VIII. APPENDIX A

The Appendix contains an overview of the all 32 ear canals geometries, ear mould geometries, and measured and propagated impedances.

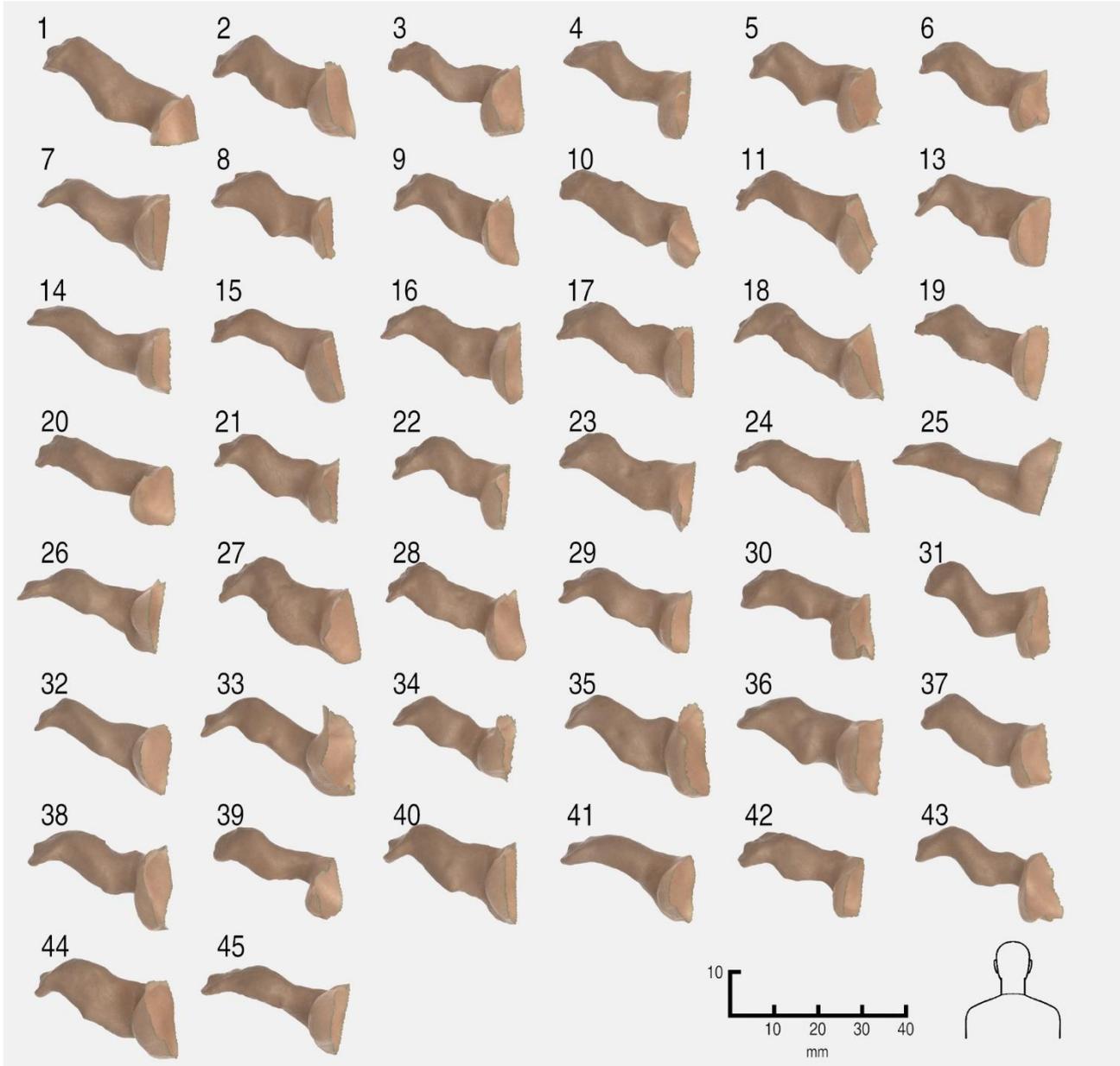

Figure A1: An overview of the 44 human ear canal geometries obtained from the MRI-based study of the human outer ear, 32 of which are the subjects included in this study. Right ear canals are shown, and the orientation corresponds to the ear canal in a person seen from the rear as indicated in the lower right part.

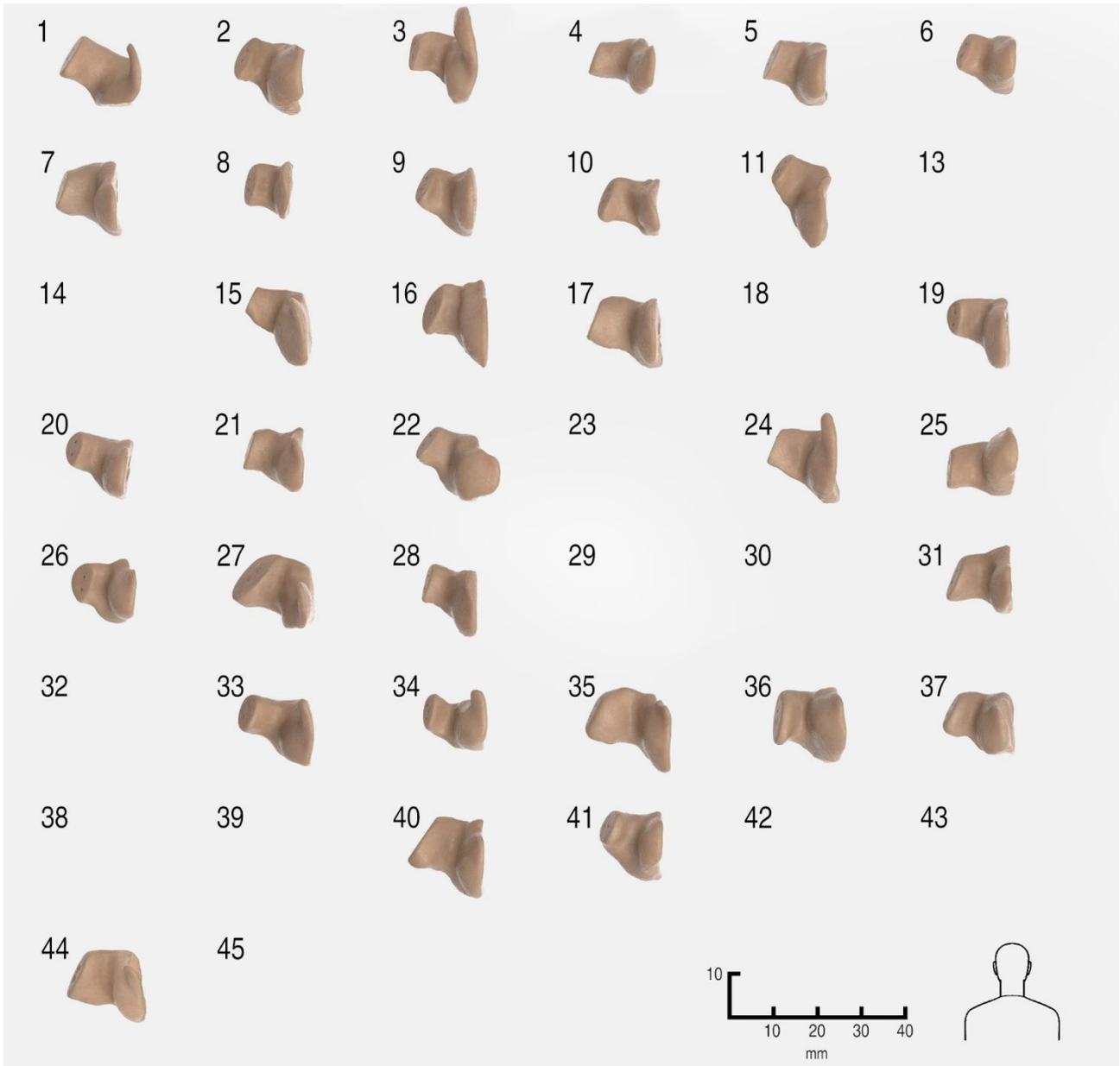

Figure A2: An overview of the 32 individual ear mould geometries for the subjects included in this study and obtained from the MRI-based study of the human outer ear. Right ear moulds are shown and the orientation corresponds to the ear canal in a person seen from the rear as indicated in the lower right part.

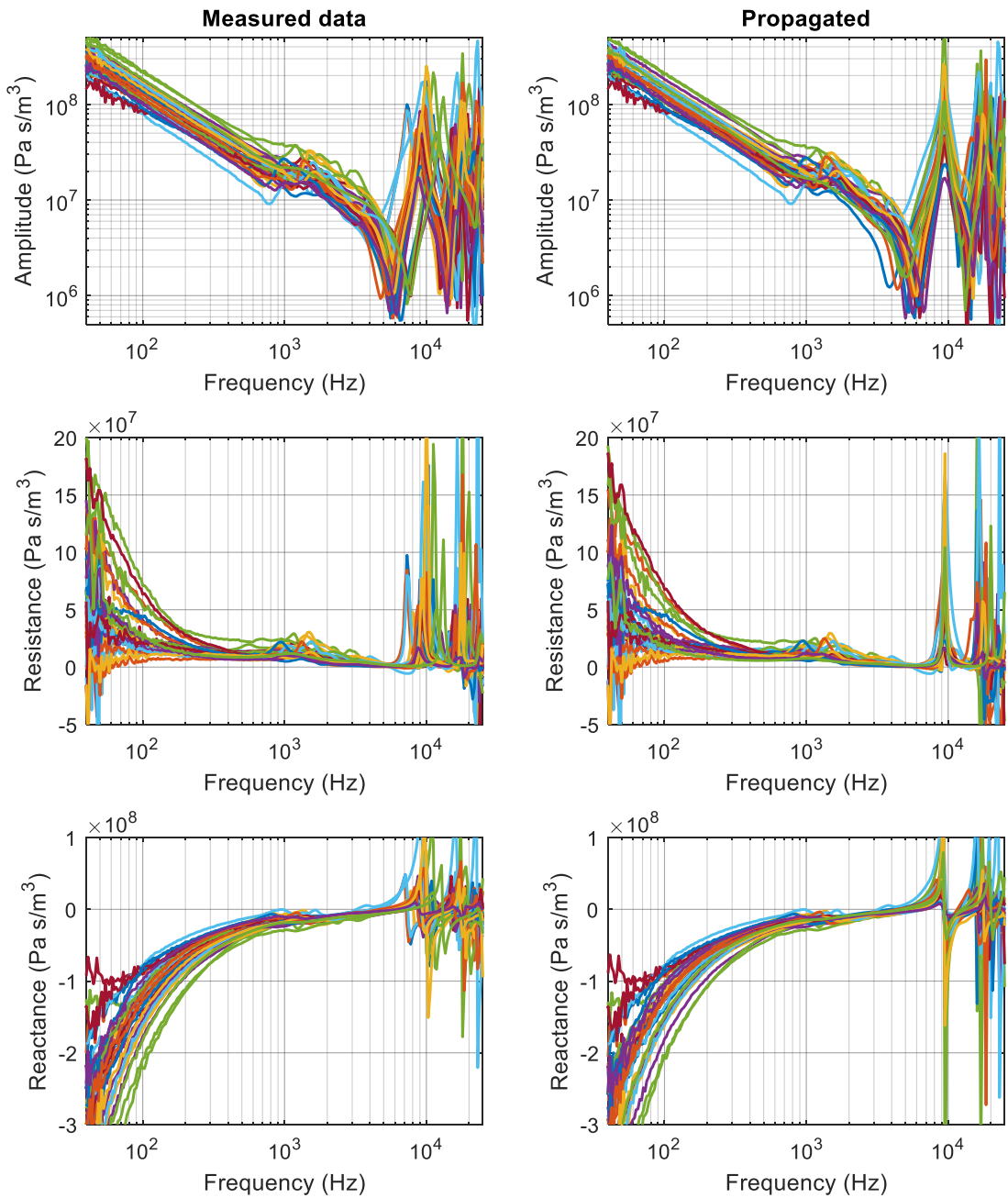

Figure A3: Measured (left column) and propagated impedance (right column) of all 32 adult subjects. The upper graphs show the amplitude and the two lower graphs show the real and imaginary part, i.e. the impedance resistance and reactance respectively.

## IX.  APPENDIX B

This appendix contains derivation of the Webster's horn propagation equation of equation ( 13 ).

In velocity potential form Webster's horn equation reads [6], [16], [17]:

$$\Phi'' + \frac{1}{A}A'\Phi' - \frac{1}{c^2}\ddot{\Phi} = 0 \qquad (b14)$$

Where $A$ is cross section area, $c$ is speed of sound and differentiation with respect to spatial coordinate $x$ and time $t$ is denoted with $'$ and $\cdot$ respectively. By using a time dependent solution of the form $\Phi = \Phi_1 \Phi_x(x)\exp(-i\omega t)$ with $k = \omega/c$, where $\omega$ is the angular frequency and $k$ is the wave number, we get:

$$\Phi_x'' + \frac{1}{A}A'\Phi_x' + k^2\Phi_x = 0 \qquad (b15)$$

By substituting new variables; $A = \pi a^2$, where $a$ is the cross-section radius, and $\psi = \Phi_x\sqrt{A}$, equation ( b14 ) reduces to a one-dimensional Schrödinger equation [18], [19]:

$$\psi'' + \frac{a''}{a}\psi = -k^2\psi \qquad (b16)$$

For a cone $a'' = 0$ and equation ( b16 ) reduces to:

$$\psi'' = -k^2\psi \qquad (b17)$$

It has solutions of the form:

$$\psi = \psi_0 \exp(\pm ikx) \qquad (b18)$$

And the solution for the velocity potential is:

$$\Phi = \frac{\Phi_1 \psi_0}{\sqrt{\pi}}\frac{1}{a}\exp(\pm ikx - i\omega t) \qquad (b19)$$

By defining a new constant $\Phi_0 = \Phi_1 \psi_0 / \sqrt{\pi}$ we get,

$$\Phi = \Phi_0 \frac{1}{a} \exp(\pm ikx - i\omega t) \quad (b20)$$

where the plus sign refers to a forward mowing wave and the minus sign refers to a backward mowing wave. Remember that for a cone the cross-section radius $a = a(x)$ depends linearly on the spatial coordinate x. The solution for a forward moving wave and a backward moving wave reflected at $x = L$ with reflection coefficient $R$ is therefore given by:

$$\Phi = \frac{\Phi_0}{a} (\exp(ikx - i\omega t) + R \exp(-ik(x + 2L) - i\omega t)) \quad (b21)$$

The sound pressure defined by $p = -\rho \frac{d\Phi}{dt}$ is given by:

$$p = i\rho\omega\Phi \quad (b22)$$

And the one-dimensional volume velocity defined by $q = A \frac{d\Phi}{dx}$ is given by,

$$q = A \left( \frac{\Phi_0}{a} (ik \exp(ikx - i\omega t) - ikR \exp(-ik(x + 2L) - i\omega t)) - \frac{a'}{a} \Phi \right) \quad (b23)$$

where $(a^{-1})' = -a'/a^2$ is used.

The acoustic input impedance defined by $Z = p/q$ can be found from equation ( 6 ) and ( 7 ):

$$Z = \left[ \frac{A \left( \frac{\Phi_0}{a} (ik \exp(ikx - i\omega t) - ikR \exp(-ik(x + 2L) - i\omega t)) - \frac{a'}{a} \Phi \right)}{i\rho\omega\Phi} \right]^{-1} \quad (b24)$$

Reducing the second term and inserting equation ( b21 ) in equation ( b24 ) yields:

$$Z = \left[ \frac{\pi a \Phi_0 ik (\exp(ikx - i\omega t) - R \exp(-ik(x + 2L) - i\omega t))}{i\rho\omega \frac{\Phi_0}{a} (\exp(ikx - i\omega t) + R \exp(-ik(x + 2L) - i\omega t))} + \frac{\pi a a'}{i\rho\omega} \right]^{-1} \quad (b25)$$

By defining the characteristic impedance $Z_0 = \rho c / A$ equation ( b25 ) can be reduced to:

$$Z = Z_0 \left[ \frac{(1 - R\exp(-2ik(x + L)))}{(1 - R\exp(-2ik(x + L)))} - \frac{ia'}{ak} \right]^{-1} \quad (b26)$$

At $x = 0$ the cross-section area is $a1$, and the impedance denoted $Z1$ is given by:

$$Z1 = \frac{\rho c}{\pi a1^2} \left[ \frac{(1 - R\exp(-2ikL))}{(1 - R\exp(-2ikL))} - \frac{ia1'}{a1 k} \right]^{-1} \quad (b27)$$

From equation ( b27 ) the reflectance times the exponential function can be found:

$$R\exp(-2ikL) = \frac{Z1\pi a1^2 ki + Z1\pi a1 a1' - \rho i\omega}{Z1\pi a1^2 ki - Z1\pi a1 a1' + \rho i\omega} \quad (b28)$$

At $x = \Delta L$ the cross-section area is $a2$, and the impedance denoted $Z2$ is given by:

$$Z2 = \frac{\rho c}{\pi a2^2} \left[ \frac{(1 - R\exp(-2ik(\Delta L + L)))}{(1 - R\exp(-2ik(\Delta L + L)))} - \frac{ia2'}{a2 k} \right]^{-1} \quad (b29)$$

Inserting equation ( b28 ) in equation ( b29 ) yields:

$$Z2 = \frac{\rho c}{\pi a2^2} \left[ \frac{\left(1 - \exp(-2ik\Delta L)\dfrac{Z1\pi a1^2 ki + Z1\pi a1 a1' - \rho i\omega}{Z1\pi a1^2 ki - Z1\pi a1 a1' + \rho i\omega}\right)}{\left(1 - \exp(-2ik\Delta L)\dfrac{Z1\pi a1^2 ki + Z1\pi a1 a1' - \rho i\omega}{Z1\pi a1^2 ki - Z1\pi a1 a1' + \rho i\omega}\right)} - \frac{ia2'}{a2 k} \right]^{-1} \quad (b30)$$

By using the trigonometric relations, $\tanh(ix) = i\tan(x)$, and $\exp(-ix) + \exp(ix) = 2\cos(x)$,

and $\tanh(x) = \frac{\exp(x) - \exp(-x)}{\exp(x) + \exp(-x)}$, equation ( b30 ) can be reduced to,

$$Z2 = Z2_0 \left[ \frac{\left(Z1_0 + i\frac{Z1a1'}{ka1}\right) - Z1i\tan(k\Delta L)}{Z1 - \left(Z1_0 + i\frac{Z1a1'}{ka1}\right)i\tan(k\Delta L)} - \frac{ia2'}{ka2} \right]^{-1} \quad (\text{b31})$$

where $Z1_0 = \rho c/A1$ and $Z2_0 = \rho c/A2$ are the characteristic impedances at the two cross-sections.